\documentclass[
  aps,
  pra,
  amsmath,amssymb,
  reprint,
]{revtex4-1}
\usepackage{graphicx,color}
\usepackage{amsmath}
\usepackage{epsfig}

\begin{document}

\title{Simple dispersion relations for Coulomb and Yukawa fluids}

\author{Sergey Khrapak and Alexey Khrapak}

\affiliation{Aix-Marseille-Universit\'{e}, CNRS, Laboratoire PIIM, UMR 7345, 13397 Marseille cedex 20, France; Institut f\"ur Materialphysik im Weltraum, Deutsches Zentrum f\"ur Luft- und Raumfahrt (DLR), Oberpfaffenhofen, Germany; Joint Institute for High Temperatures, Russian Academy of Sciences, Moscow, Russia}

\date{\today}

\begin{abstract}
Very simple explicit analytical expressions, which are able to describe the dispersion relations of collective modes in strongly coupled plasma fluids, are summarized. The accuracy of these expressions is demonstrated using the comparison with available results from benchmark numerical simulations.   
\end{abstract}

\maketitle

\section{Introduction}

Transport and collective dynamics in strongly coupled plasmas is an important current research topic~\cite{Merlino2014,StricklerPRX2016} with clear interdisciplinary relations (e.g. to collective motion in other condensed matter systems)~\cite{KhrapakJCP2016,KhrapakSciRep}. It is particularly relevant for complex (dusty) plasmas, representing small charged particles immersed in the plasma medium~\cite{FortovUFN,FortovPR,Bonitz2010}, because these are normally found in strongly coupled regime (due to strong electrical interactions between the particles) and naturally form condensed liquid and solid phases. 

A very useful theoretical approach to describe waves in strongly coupled plasmas is known as the quasilocalized charge approximation (QLCA)~\cite{GoldenPoP2000,RosenbergPRE1997,KalmanPRL2000,DonkoJPCM2008}. The QLCA is based on the physical model that suggests that for a substantial portion of their time history particles in a strongly interacting system
are trapped and oscillate in the momentary local minima
of the fluctuating potential (the so-called caging effect)~\cite{KalmanPRL2000}. In fact, the generic expressions of the QLCA approach can be traced back to the calculations of high-frequency elastic moduli and quasi-crystalline approximation (QCA) to neutral liquids back in 1960s~\cite{Zwanzig1965,Zwanzig1967,Schofield1966,Nossal1968,Hubbard1969,
Takeno1971}. Also, the same equations for the contribution from potential interactions between the particles come from the analysis of the fourth and second frequency moments of the dynamical structure factor~\cite{deGennes1959,HansenBook}. In the following we call this class of theoretical approximations as the QCA/QLCA formalism.   

Within the QLCA/QCA formalism the wave dispersion relations are explicitly expressed in terms of the pairwise interparticle interaction potential $V(r)$ and the equilibrium radial distribution function (RDF) $g(r)$, characterizing structural properties of the system. The generic expressions are very simple~\cite{Hubbard1969,Takeno1971}:
\begin{equation}\label{w_L}
\omega_{\rm L}^2=\frac{n}{m}\int\frac{\partial^2 V(r)}{\partial z^2} g(r) \left[1-\cos(kz)\right]d{\bf r},
\end{equation} 
\begin{equation}\label{w_T}
\omega_{\rm T}^2=\frac{n}{m}\int\frac{\partial^2 V(r)}{\partial x^2} g(r) \left[1-\cos(kz)\right]d{\bf r},
\end{equation}
where $n$ is the particle number density, $m$ is the particle mass, $\omega$ is the frequency, $k$ is the wave number, and $z=r\cos\theta$ is the direction of the propagation of the longitudinal wave. Here $\omega_{\rm L}$ stands for the frequency of the longitudinal mode and $\omega_{\rm T}$ for that of the transverse mode. 

As displayed in Eqs.~(\ref{w_L}) and (\ref{w_T}), the dispersion relations within QCA/QLCA formalism are completely determined by the potential interaction between the particles. In related approaches (e.g. based on sum rules or generalized high-frequency elastic moduli) kinetic terms can naturally appear additively to the potential terms ($3k^2v_T^2$ for the longitudinal mode and $k^2v_T^2$ for the transverse mode, where $v_{\rm T}=\sqrt{T/m}$ is the thermal velocity)~\cite{HansenBook,Zwanzig1967,Nossal1968}. Shofield argued that these kinetic terms should not be included in the dispersion relation~\cite{Schofield1966}. However, independently of whether included or not, the kinetic contribution is numerically small at liquid densities (i.e. at strong coupling). Since we focus on strongly coupled regime in this manuscript, in the following we will use dispersion relations ~(\ref{w_L}) and (\ref{w_T}), completely neglecting kinetic contributions.

The radial distribution function  $g(r)$, appearing under the integrals in Eqs.~(\ref{w_L}) and (\ref{w_T}) can in principle be obtained from direct numerical simulations or from appropriate integral equations from liquid state theory. Nowadays, there are no principle difficulties in obtaining RDFs for a given interaction potential and at a required phase state point. Then, the integration in Eqs.~(\ref{w_L}) and (\ref{w_T}) can be performed numerically, which will yield the longitudinal and transverse dispersion relations for a given particle configuration. It has been recently shown, however, that to describe the long-wavelength portions of the dispersion curves a very accurate knowledge of $g(r)$ is unnecessary~\cite{KhrapakPoP2016}. The main idea is that since in the QCA/QLCA model the function $g(r)$ appears under the integral, an appropriate models for $g(r)$, even if it does not describe very accurately the actual structural properties of the system, can nevertheless be helpful in estimating the behaviour of dispersion curves. 
 
A very simple model form of $g(r)$, appropriate for soft repulsive interparticle interactions operating in plasma-related context has been proposed in Refs.~\cite{KhrapakPoP2016,2DOCP,KhrapakEPL2017}. This RDF reads
\begin{equation}\label{gOFr}
 g(r)=\theta(r-R),
\end{equation}
where $\theta(x)$ is the Heaviside step function and $R$ is the distance of order of the mean interparticle separation. Physically, this trial form seems reasonable, because the main contribution to the long-wavelength dispersion corresponds to long length scales, where $g(r)\simeq 1$. The excluded volume effect for $r\leq R$ allows us to properly account for strong coupling effects. Earlier, a similar RDF was employed to analyse the main tendencies in the behaviour of specific heat of liquids and dense gases at low temperatures~\cite{Stishov}. The radius $R$ is generally not a free parameter of the approximation~\cite{KhrapakPoP2016}, but can be determined from the condition that the model form (\ref{gOFr}) delivers good accuracy for the excess internal energy and pressure, which can also be expressed as integrals over $g(r)$ for pairwise interactions~\cite{HansenBook,MarchBook}. A simple analytical expression for the relation between the excluded cavity radius $R$ and the coupling strength in Yukawa systems has been recently derived~\cite{KhrapakEPL2017}.
Combined with the fact that for the model RDF (\ref{gOFr}), Eqs. (\ref{w_L}) and (\ref{w_T}) can be integrated analytically, the procedure results in very simple dispersion relations for strongly coupled plasma-related systems (one-component Yukawa systems along with the limiting case of the Coulomb one-component plasma). These fully analytical dispersion relations demonstrate high accuracy in the long-wavelength limit. Overall, tremendous simplification of the description of the waves dispersion relations in strongly coupled plasma-related systems has been reached.    

The purpose of the present paper is to give a short summary of the approach discussed above along with a few related recent results. We provide new comparison, which documents again the success of the proposed approximation in the case of longitudinal mode. Regarding the dispersion relation of the transverse mode, we propose a simple procedure to account for the disappearance of the shear mode at $k\rightarrow 0$ and the existence of the corresponding cutoff wave-vector $k_*$, which are well known properties of the liquid state, but cannot be accounted for within the original QCA/QLCA formalism. Overall, we report a tool which allows very simple and accurate description of collective modes in strongly-coupled plasma fluids.    

\section{Approach}  

In the following we consider Yukawa systems, characterized by the repulsive interaction potential of the form $V(r)=(Q^2/r)\exp(-r/\lambda)$, where $Q$ is the particle charge, $\lambda$ is the screening length, and $r$ is the distance between a pair of particles. The phase state of the system is conventionally described  by the two dimensionless parameters~\cite{HamaguchiPRE1997}, which are the coupling parameter $\Gamma=Q^2/aT$ and the screening parameter $\kappa=a/\lambda$. Here $T$ is the system temperature (in energy units) and $a=(4\pi n/3)^{-1/3}$ is the Wigner-Seitz radius (three-dimensional systems are considered here). When $\kappa\rightarrow 0$ the one-component-plasma (OCP) limit is recovered, however neutralizing background should be added to keep thermodynamic quantities finite~\cite{Baus1980}. Yukawa systems are normally referred to as strongly coupled when $\Gamma\gg 1$. When $\Gamma$ reaches sufficiently high values, the fluid-solid (crystallization) transition takes place at $\Gamma=\Gamma_{\rm m}(\kappa)$~\cite{HamaguchiPRE1997,VaulinaJETP2000,VaulinaPRE2002}. A typical range for screening parameter in experiments with three-dimensional complex plasmas normally corresponds to $2\lesssim \kappa \lesssim 6$~\cite{KhrapakPRL2011,KhrapakPRE2012}. The Yukawa potential is considered as a reasonable starting point (zero approximation) to model interactions in complex (dusty) plasmas and colloidal dispersions~\cite{FortovPR,IvlevBook,KhrapakCPP2009,ChaudhuriIEEE2010}. In this paper we consider an idealized Yukawa fluid and neglect all types of collisions other than particle-particle electrical interactions. In the context of complex plasmas collisions between charged particles and neutral atoms or molecules are present and usually represent an important mechanism of damping. This neutral damping effect can be relatively easily included in the QLCA formalism in an {\it ad hoc} manner~\cite{RosenbergPRE1997,KalmanPRL2000,Fingerprints} and we will not consider this further.

For the Yukawa interaction potential and RDF model of Eq. (\ref{gOFr}), the generic expressions for the QLCA dispersion relations (containing rather complex integrals~\cite{KalmanPRL2000,DonkoJPCM2008,Fingerprints}) can be integrated analytically. This results in simple and elegant expressions for the frequencies of the longitudinal and transverse modes~\cite{KhrapakPoP2016}:
\begin{equation}\label{L1}
\begin{aligned}
\omega_{\rm L}^2=\omega_{\rm p}^2e^{-R\kappa}\left[\left(1+R\kappa\right)\left(\frac{1}{3}-\frac{2\cos Rq}{R^2q^2}+\frac{2\sin Rq}{R^3q^3} \right) \right. \\ \left. -\frac{\kappa^2}{\kappa^2+q^2}\left(\cos Rq+\frac{\kappa}{q}\sin Rq \right)\right],
\end{aligned}
\end{equation}  
and
\begin{equation}\label{T1}
\omega_T^2=\omega_{\rm p}^2e^{-R\kappa}\left(1+R\kappa\right)\left(\frac{1}{3}+\frac{\cos Rq}{R^2q^2}-\frac{\sin Rq}{R^3q^3} \right).
\end{equation}
Here $q=ka$ is the reduced wave number, $\omega_{\rm p}=\sqrt{4\pi Q^2 n/m}$ is the plasma frequency scale and $R$ is expressed in units of $a$. 

The remaining step is to determine the radius $R$ of the excluded volume cavity that should be used in calculations. A simple analytical expression has been recently suggested in Ref.~\cite{KhrapakEPL2017} by equating the excess energy corresponding to the model RDF of Eq.~(\ref{gOFr}) and the excess energy of the ion sphere model (ISM)~\cite{KhrapakISM,KhrapakPRE2015_1,KhrapakJCP2015}. The result is  
\begin{equation}\label{R}
R(\kappa)\simeq 1+\frac{1}{\kappa}\ln \left[\frac{3 \cosh (\kappa)}{\kappa^2}-\frac{3 \sinh
(\kappa)}{\kappa^3}\right].
\end{equation}
Note that this can be further simplified to $R(\kappa)\simeq 1+\kappa/10$ in the regime of sufficiently weak screening. The excluded volume radius turns out to be independent of $\Gamma$, because only the dominant  (static) contribution to the excess energy is taken into account. The static excess energy per particle, normalized to the system temperature, is directly proportional to $\Gamma$~\cite{KhrapakISM,KhrapakPRE2015_1}, so $\Gamma$ simply cancels out when evaluating $R$.    

The set of expressions (\ref{L1}) and (\ref{T1}) combined with Eq.~(\ref{R}) represents a very simple practical tool to describe the dispersion relations of longitudinal and transverse modes in strongly coupled Yukawa fluids. Below, we demonstrate the high accuracy of this approximation by comparing it with the results from benchmark numerical simulations. Before we do that, let us just briefly summarize some recent results corresponding to the long-wavelength limit.   

In the long-wavelength limit, Eqs.~(\ref{L1}) and (\ref{T1}) yield acoustic dispersion relations and the longitudinal and transverse sound velocities can be introduced,
\begin{equation}
\lim_{k\rightarrow 0} \frac{\omega_{\rm L/T}^2}{k^2} = C_{\rm L/T}^2.
\end{equation}      
Within the discussed approach we easily obtain
\begin{equation}\label{SoundL}
c_{\rm L}^2=\frac{\exp(-\kappa R)}{\kappa^2}\left(1+\kappa R+\tfrac{13}{30}\kappa^2 R^2 + \tfrac{1}{10}\kappa^3 R^3\right),
\end{equation}
and 
\begin{equation}\label{SoundT}
c_{\rm T}^2=\frac{\exp(-\kappa R)}{30}R^2\left(1+\kappa R\right),
\end{equation}
where the reduced velocities $c_{\rm L}$ and $c_{\rm T}$ are expressed in units of $\omega_{\rm p}a$. (Note that $C$ and $c$ denote the dimensional and dimensionless sound velocities throughout the paper). To get the sound velocity in units of thermal velocity, $v_{\rm T}=\sqrt{T/m}$, one should simply multiply $c_{\rm L}$ and $c_{\rm T}$ by the factor $\sqrt{3\Gamma}$. 

Two useful relations between $c_{\rm L}$ and $c_{\rm T}$ have been recently discussed (for more details see Ref.~\cite{KhrapakPoP2016_Relations}).
First, $c_{\rm L}$ and $c_{\rm T}$ are related to the reduced excess pressure $p_{\rm ex}$ via
\begin{equation}
c_{\rm L}^2-3c_{\rm T}^2=\frac{2p_{\rm ex}}{3\Gamma}.
\end{equation}    
This is a general (independent of the interaction potential) property of the QCA model in both two- and three dimensions~\cite{KhrapakPoP2016_Relations}. In addition, the longitudinal and transverse sound velocities are related to the instantaneous sound velocity $c_{\infty}$ via~\cite{KhrapakSciRep,Schofield1966} 
\begin{equation}
c_{\infty}^2=c_{\rm L}^2-\tfrac{4}{3}c_{\rm T}^2.
\end{equation} 
This is very similar to the conventional relations for elastic waves in an isotropic medium~\cite{LL}. The instantaneous sound velocity $c_{\infty}$ appears to be rather close to the conventional thermodynamic (adiabatic) sound velocity $c_{\rm s}$~\cite{KhrapakPRE2015_Sound,KhrapakPPCF2016} for soft Yukawa systems considered here. This has been recently documented~\cite{KhrapakPoP2016_Relations}. Moreover, for such soft interactions the strong inequality $c_{\rm L}\gg c_{\rm T}$ normally holds, which implies $c_{\rm L}\simeq c_{\infty}\simeq c_{\rm s}$.  However, this is not the case for steeper interactions. In particular, the QCA/QLCA approach itself breaks down on approaching the limit of the hard sphere interactions~\cite{KhrapakSciRep}.

\section{Validation}

\begin{figure}
\includegraphics[width=8cm]{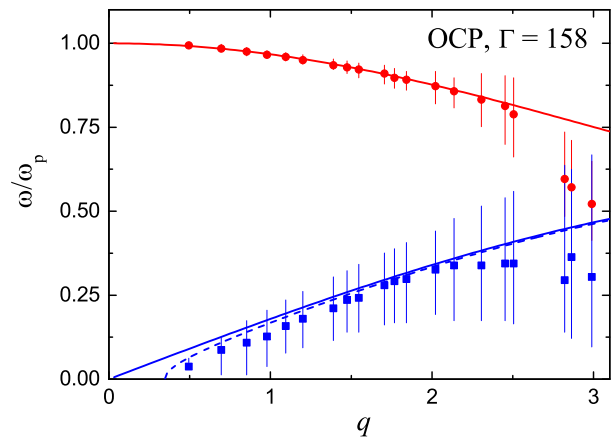}
\caption{Dispersion of the longitudinal (red) and transverse (blue) modes of the strongly coupled  OCP with $\Gamma= 158$. Symbols correspond to the results from MD simulations by Schmidt {\it et al}.~\cite{SchmidtPRE1997} The solid curves are plotted using Eq.~(\ref{OCPdispL}) and ~(\ref{OCPdispT}) with  $R=1.0$. The dashed curve displays the modification to the dispersion relation of the transverse mode discussed in the text.}
\label{Fig1}
\end{figure}

\begin{figure*}
\centering
\includegraphics[width=16cm]{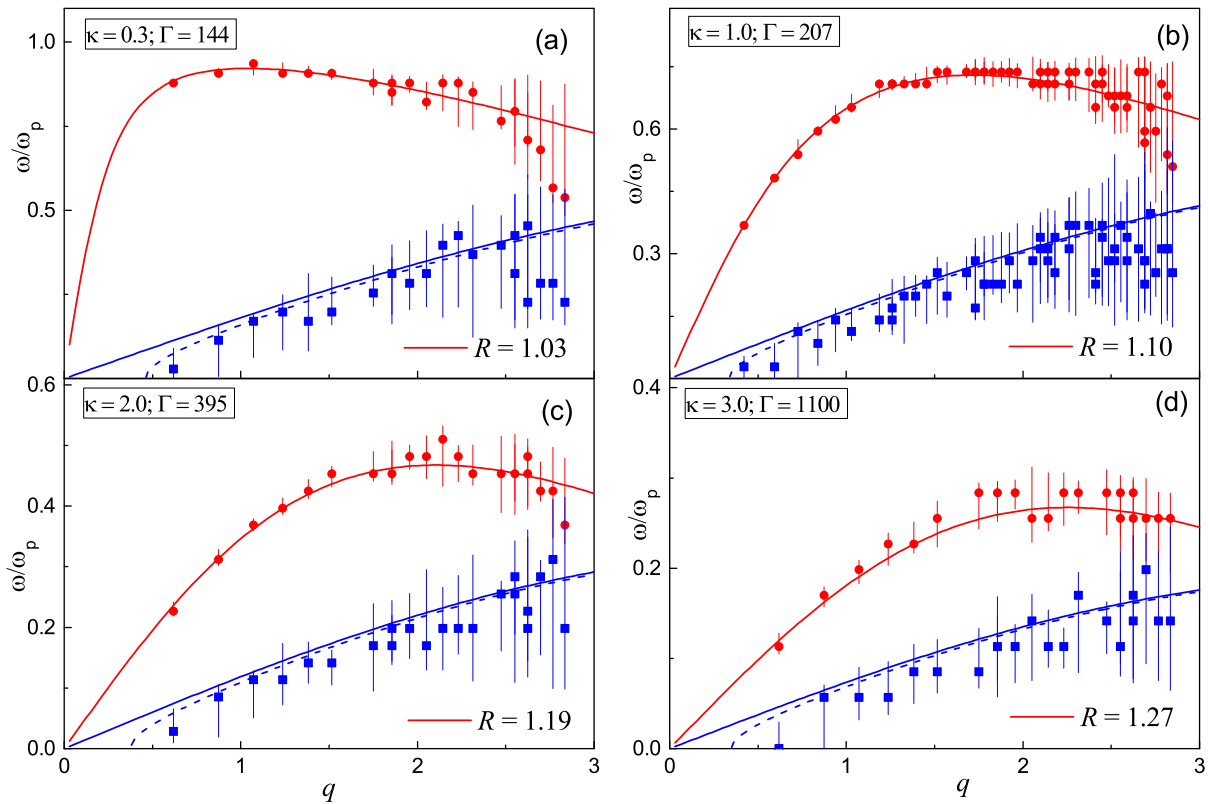}
\caption{Dispersion of the longitudinal (red) and transverse (blue) waves in Yukawa fluids near the fluid-solid phase transition (the values of $\kappa$ and $\Gamma$ are indicated in the top left corners of Figs.~(a)-(d)). Symbols with error bars correspond to the results from numerical simulations~\cite{OhtaPRL2000,HamaguchiPS2001}. Solid curves are calculated using Eqs.~(\ref{L1}) and (\ref{T1}) along with the determination of $R$ in Eq.~(\ref{R}). The corresponding values of $R$ appear in the bottom right corners (note, that $R\simeq 1+\kappa/10$ holds to a good accuracy). Dashed curves correspond to the dispersion relations of the transverse modes taking into account an {\it ad hoc} modification proposed in this work. The values of $1/4\tau_{\rm R}^2\omega_{\rm p}^2$ were estimated using the procedure similar to that used in the case of OCP, with the values of $\eta_*$ taken from Refs.~\cite{DonkoPRE2008,DaligaultPRE2014}.}
\label{Fig2}
\end{figure*}

We start with the OCP limit, corresponding to the unscreened Coulomb interaction between the particles. The dispersion relation of the longitudinal plasmon mode follows directly from Eq.~(\ref{L1}) by taking the limit $\kappa\rightarrow 0$:
\begin{equation}\label{OCPdispL}
\omega_{\rm L}^2=\omega_{\rm p}^2\left(\frac{1}{3}-\frac{2\cos Rq}{R^2q^2}+\frac{2\sin Rq}{R^3q^3} \right).
\end{equation}
Similarly, the dispersion relation of the transverse mode is
\begin{equation}\label{OCPdispT}
\omega_{\rm T}^2=\omega_{\rm p}^2\left(\frac{1}{3}+\frac{\cos Rq}{R^2q^2}-\frac{\sin Rq}{R^3q^3} \right).
\end{equation}   
Note that the Kohn sum rule is automatically satisfied, $\omega_{\rm L}^2+2\omega_{\rm T}^2=\omega_{\rm p}^2$. An approximate equation (\ref{R}) yields $R=1$ in this limit. A somewhat more accurate analysis, which takes into account specifics of the OCP, results in a rather close value of $R= \sqrt{6/5}\simeq 1.09545$~\cite{KhrapakPoP2016}, which is, however, not used here.

The longitudeinal and transverse dispersion relations calculated with the help of Eqs.~(\ref{OCPdispL}) and (\ref{OCPdispT}) are plotted in Fig.~\ref{Fig1} for a strongly coupled OCP state point with $\Gamma = 158$. The symbols correspond to the results derived from MD simulations by Schmidt {\it et al.}~\cite{SchmidtPRE1997}. The agreement is very good, except the initial (small $q$) portion of the transverse dispersion relation. 

We should remind that the disappearance of the shear mode at $q\rightarrow 0$ and the existence of the corresponding cutoff wave-vector $q_*$	are well known properties of the liquid state, which cannot be properly described within the conventional QCA/QLCA formalism, because the damping effects are not included. We propose the following simple {\it ad hoc} procedure to improve the situation. In the long-wavelength limit we expect from the generalized hydrodynamic consideration the transverse mode dispersion of the form~\cite{OhtaPRL2000,Yang2017}
\begin{equation}
\omega_{\rm T}^2\simeq \omega_{\rm p}^2 c_{\rm T}^2q^2-\frac{1}{4\tau_{\rm R}^2},
\end{equation}
where $\tau_{\rm R}$ is the relaxation time. The proposed modification is to simply add the term $-1/4\tau_{\rm R}^2$ to the left-hand-side of the dispersion relation (\ref{OCPdispT}), or (\ref{T1}) in the case of Yukawa systems. In this way, the dispersion relation becomes correct in the long-wavelength low-frequency regime. In the high-frequency regime, where $\omega \tau_{\rm R}\gg 1$, this correction is negligible.

The relaxation time appears as a coefficient of proportionality between the shear viscosity $\eta$ and the shear modulus $G$, $\eta \simeq G \tau_{\rm R}$ (this is often called the Maxwellian relationship). In addition, the shear modulus is related to the transverse sound velocity via $C_{\rm T}^2=G/mn$.
Combining all this, we get 
\begin{equation}
\tau_{\rm R}\omega_{\rm p} = \eta_*^2/c_{\rm T}^4,
\end{equation}
where the reduced shear viscosity is $\eta_*=\eta/(mn\omega_{\rm p}a^2)$. The values of $\eta_*$ have been repeatedly tabulated in literature, see e.g. Refs.~\cite{DonkoPRE2008,DaligaultPRE2014}. The values of $c_{\rm T}$ can be estimated directly from Eq.~(\ref{SoundT}). In doing so we get $1/4\tau_{\rm R}^2\omega_{\rm p}^2\simeq 0.004$ for the strongly coupled OCP at $\Gamma \sim 160$. The resulting transverse wave dispersion is shown in Fig.~\ref{Fig1} by the dashed curve. The modification allows to describe the existence and (roughly) the amplitude of the cutoff wave vector $q_*$, although the agreement with numerical results is not perfect.         

It should be mentioned that the level of accuracy documented in Fig.~\ref{Fig1} for the longitudinal mode should not be expected for weaker coupling. This is because the original QLCA model is not designed for this regime~\cite{GoldenPoP2000} and hence lacks accuracy.  In particular, it cannot describe the transition from the positive to negative dispersion~\cite{MithenAIP2012,HansenJPL1981} at $\Gamma\simeq 10$. Here positive/negative dispersion refers to the positive/negative sign of $d\omega/d q$ at $q\rightarrow 0$ (note, that in this sense the dispersion is always negative within the original QCA/QLCA formalism). Useful modifications which allow to capture the onset of negative dispersion at moderate coupling have been recently discussed~\cite{KhrapakNegative}.  

Finally, we compare the results of calculations using Eqs.~(\ref{L1}) -- (\ref{R}) with the results obtained for the dispersion of Yukawa fluids using molecular dynamics simulations by Hamaguchi and Ohta~\cite{OhtaPRL2000,HamaguchiPS2001}. This comparison is shown in Fig.~\ref{Fig2} for four state points, located near the fluid-solid (freezing) curve. In all cases the agreement is impressive, especially taking into account the level of simplifications involved. The proposed modification to the transverse dispersion relation demonstrates considerable improvement over the conventional QCA/QLCA approximation, although some room for further improvement clearly remains.

\section{Summary}

The main results reported above can be summarized as follows. Simple approximation for the radial distribution function $g(r)$ has been used to derive analytical expressions for the dispersion of longitudinal and transverse modes in strongly coupled plasma fluids, based on the QCA/QLCA model. The accuracy of this approximation has been tested against the benchmark results from previous numerical simulations and a very good agreement has been documented. A simple procedure to improve the QCA/QLCA performance near the cutoff wave vector of the transverse mode has been discussed. Most of the results reported are unlikely limited to the particular shape of the interaction potentials (Yukawa and Coulomb) considered here, but should also apply to other classical particle systems with sufficiently soft interactions. This opens up exciting perspectives for  interdisciplinary application of these results.

\section*{Acknowledgments}

We would like to thank our colleagues: B. Klumov, L. Couedel, and H. Thomas who participated in different studies related to the content of this paper. We also thank R. Kompaneets for careful reading of the manuscript. The studies described in this paper were supported by the A*MIDEX project (Nr.~ANR-11-IDEX-0001-02) funded by the French Government ``Investissements d'Avenir'' program managed by the French National Research Agency (ANR) and by the Russian Science Foundation (grant RSF 14-12-01235).

\bibliographystyle{aipnum4-1}
\bibliography{IEEE_References}

\end{document}